\documentstyle[11pt]{article}

\begin{document}

\begin{center}
{\Large Radiation from an electron bunch\\[0pt]
flying over a surface wave}

\bigskip

A. A. Saharian\footnote{%
E-mail: saharyan@www.physdep.r.am}, A. R. Mkrtchyan\footnote{%
E-mail: malpic@iapp.sci.am}, L. A. Gevorgian,

L. Sh. Grigorian\footnote{%
E-mail: levonshg@iapp.sci.am}, B. V. Khachatryan,

\bigskip

{\em Institute of Applied Problems in Physics \\[0pt]
25 Nersessian Str., 375014 Yerevan, Armenia}
\end{center}

\bigskip

\noindent
{\bf Abstract.} Radiation generated by the passage of a monoenergetic
electron bunch above the surface wave excited in plane interface between
homogeneous media with different dielectric constants is investigated. For
the surface wave of general profile the radiation intensity is expressed via
the radiated power from a single charge and bunch form factor. Various types
of transverse and longitudinal distributions of electrons in the bunch have
been considered including Gaussian, asymmetrical Gaussian, two Gaussian and
rectangular distribution with asymmetrical exponential tails. Conditions are
specified under which the coherent radiation essentially exceeds the
incoherent part.

\bigskip


\bigskip

\section{Introduction}

Recently a great deal of attention has been devoted to the investigations of
the radiation from charged particles in periodic structures. Such radiation
has a number of remarkable properties and is widely used in various regions
of science and technology (see, for example, \cite{bolot} - \cite{rreps},
and references therein). The possible physical applications include the
generation of the electromagnetic radiation in various wavelength ranges by
beams of charged particles and the determination of the characteristics of
emitting particles by using the properties of the radiation field. As an
example one can mention here the transition radiation from a charge
traversing a stack of plates or moving in a medium with periodically varying
dielectric permittivity \cite{Termik}, \cite{Garib}, \cite{Ginz} . Another
important example is the Smith-Purcell radiation, which arises when charged
particles are in flight near a diffraction grating (see e.g.\cite{bolot,
beam, shest, wort}). This radiation is one of the main mechanisms for the
generation of electromagnetic waves in the millimeter and submillimeter
wavelength range (see \cite{shest}).

Recent theoretical and experimental developments \cite{gab}-\cite{grig3}
have shown that the use of external (ultrasonic, temperature gradient)
fields have a potential to control various parameters of emitted radiation.
The modification of the crystal lattice by external fields can result in
radiation intensity enhancement. In papers \cite{mkrt1}-\cite{mkrt3} we have
considered the radiation from a charge flying over a surface acoustic wave
excited on the plane interface between homogeneous media with different
dielectric constants. It is shown that the use of surface waves essentially
simplifies the control of angular-frequency characteristics of the emitted
radiation.

The present paper deals with the consideration of the similar issue for
relativistic electron bunches. It also generalizes similar results in \cite
{mkrt4} for coherent diffraction radiation from an electron bunch. In
Section 2 we derive a formula for the radiation intensity of an electron
bunch flying over the surface wave of an arbitrary profile. In Section 3 we
consider various types of transverse and longitudinal distribution of
electrons in the bunch and specify conditions under which the coherent part
of the radiation essentially exceeds its incoherent part.

\section{Radiation intensity}

Assume that a monoenergetic bunch of $N$ particles moves at constant
velocity $v$ along the $z$ axis, parallel to the propagation direction of a
surface wave excited on plane boundary between homogeneous media with
perimittivities $\varepsilon _1$ and $\varepsilon _2$. If $x$ and $y$ axes
of the Cartesian system of coordinates are directed perpendicular and
parallel to the interface respectively, then the relation between
coordinates $x,y,z$ of the interface may be written as 
\begin{equation}
x=-d+f(\xi ),\quad \xi \equiv k_0z\mp \omega _0t  \label{profile}
\end{equation}
where $k_0$ and $\omega _0$ are the wave number and frequency of the surface
wave, $d$ is the distance of the non-excited interface from origin of
coordinate system and $f$ is a function describing the surface wave profile.
The minus/plus signs correspond to the propagation of wave in the
positive/negative direction of $z$ axis.

The radiation spectrum does not depend on surface wave profile and can be
obtained by using the following arguments. In the bunch rest frame the
radiation frequency $\nu ^{\prime }$ is multiple to the frequency of passing
one wavelength $\lambda _0^{\prime }$ of surface wave: $\nu ^{\prime
}=mv_0^{\prime }/\lambda _0^{\prime }$, $m$ is an integer, $v_0^{\prime
}=(v_0\mp v)/(1-v_0v/c^2)$ is the velocity of surface wave in the bunch rest
frame, and $v_0=\omega _0/k_0$. If $\lambda _0$ is the corresponding
wavelength in the laboratory system then $\lambda _0^{\prime }=\lambda _0%
\sqrt{1-v^2/c^2}/(1-v_0v/c^2)$. By substituting these relations into the
standard transformation formula for the frequency, one obtains the relation
between the radiation frequency $\omega $ and emission angle in the
laboratory system. For the medium with permittivity $\varepsilon _\alpha $
this relation has the form (see, for example, \cite{mkrt2}) 
\begin{equation}
\omega =\frac{m(k_0v\mp \omega _0)}{1-\beta \sqrt{\varepsilon _\alpha }\cos
\theta },\quad \beta =v/c  \label{disrel}
\end{equation}
where $\theta $ is the angle between the bunch velocity and the radiation
direction. When $\omega _0=0$ we obtain the standard relation for the static
case \cite{bolot}. Note that in the case $v\gg v_0$ one has $\omega \gg
m\omega _0$, whereas $\omega \sim m\omega _0$ for $v\sim v_0$. The relation (%
\ref{disrel}) can be derived also from the conditions of system periodicity
under the transformations 
\begin{equation}
t\rightarrow t+\frac{\lambda _0}{v\mp v_0},\quad z\rightarrow z+\frac{%
\lambda _0v}{v\mp v_0}  \label{pertrans}
\end{equation}
In particular, it follows from here that the projection of the wave vector
on the axis $z$, $k_z=\omega \sqrt{\varepsilon _\alpha }\cos \theta /c$, is
equal to 
\begin{equation}
k_{zm}=\frac \omega v-m\left( k_0\mp \frac{\omega _0}v\right)   \label{kazet}
\end{equation}
The relation (\ref{disrel}) is a direct consequence of this formula. It can
be easily seen that Eq. (\ref{kazet}) corresponds to that the waves emitted
from the two neighboring humps of the surface wave are in phase.

As seen from (\ref{disrel}) in the case of infinite media the photons
radiated in fixed direction $\theta $ have frequencies multiple to the main
harmonic with $m=1$. If the media have a finite size, $2z_p$, in direction
of bunch moving ($z$ axis) then in the formula for the radiation intensity
instead of $\delta $-function one has $\sin [z_p(k_z-k_{zm})]/\pi
(k_z-k_{zm})$ with $k_{zm}$ from Eq.(\ref{kazet}). Now one has a sharp peak
at frequency given by (\ref{disrel}) with width at half maximum equal to 
\begin{equation}
\Delta \omega =\frac{2v}{z_p(1-\beta \sqrt{\varepsilon _\alpha }\cos \theta )%
}  \label{finitesize}
\end{equation}
For $z_p\gg \lambda _0$ the distance between the neighboring harmonics is
much larger than the width of the spectral line.

The electric and magnetic fields $\vec{E}_\alpha $ and $\vec{H}_\alpha $ in
the regions with permittivity $\varepsilon _\alpha $ ($\alpha =1,2$) are
determined by the Maxwell equations with corresponding boundary conditions
on the interface (\ref{profile}). We can write the fields in the first
medium as 
\begin{equation}
\vec{E}_1=\vec{E}_0+\vec{E}_1^{^{\prime }},\qquad \vec{H}_1=\vec{H}_0+\vec{H}%
_1^{^{\prime }},  \label{fields1}
\end{equation}
where $\vec{E}_0,\vec{H}_0$ are the fields of bunch at its motion in the
homogeneous medium with permittivity $\varepsilon _1$. It is convenient to
present the expressions for these fields in the form 
\begin{eqnarray}
\vec{F}_0 &=&\sum_{j=1}^N\int d\omega dk_y\vec{F}_0(\omega ,k_y)e^{-i\vec{k}%
\vec{R_j}}e^{i(\vec{k}\vec{r}-\omega t)},\qquad F=E,H  \label{fields2} \\
\vec{E}_0\left( \omega ,k_y\right) &=&\frac q{2\pi vg_0}\left( \frac{\vec{k}%
}{\varepsilon _1}-\frac \omega {c^2}\vec{v}\right) ,\quad \vec{H}_0(\omega
,k_y)=\frac q{2\pi vg_0}\left[ \frac{\vec{v}}c\vec{k}\right]  \nonumber
\end{eqnarray}
where the vector $\vec{R}_j=(X_j,Y_j,Z_j)$ gives the position of the $j$-th
particle in the bunch at the initial moment $t=0$. In Eq. (\ref{fields2})
the factor depending on the particle number is separated and the following
notations are introduced: 
\begin{equation}
g_0=\left( \frac{\omega ^2}{c^2}\varepsilon _1-k_y^2-\frac{\omega ^2}{v^2}%
\right) ^{1/2},\quad \vec{k}=\left( g_0sgn\left( x-X_j\right) ,k_y,\omega
/v\right)  \label{notat1}
\end{equation}

The fields $\vec{E}_1^{^{\prime }},\vec{H}_1^{^{\prime }},\vec{E}_2,\vec{H}_2
$ are solutions of homogeneous Maxwell equations. With due regard for the
system periodicity these fields can be written in the form of Fourier
expansions 
\begin{equation}
\vec{F}_\alpha =\sum_{j=1}^N\sum_{m=-\infty }^{+\infty }\int d\omega dk_y%
\vec{F}_{\alpha m}^{(j)}\left( \omega ,k_y\right) e^{i\left( \vec{k}%
^{(\alpha )}\vec{r}-\omega t\right) },\quad \alpha =1,2  \label{fields3}
\end{equation}
where $k_z$ is determined by the relation (\ref{kazet}), 
\begin{equation}
\vec{k}^{\left( \alpha \right) }=\left( k_x^{\left( \alpha \right)
},k_y,k_{zm}\right) ,\qquad k_x^{\left( \alpha \right) }=\left( \frac{\omega
^2}{c^2}\varepsilon _\alpha -k_y^2-k_{zm}^2\right) ^{1/2}  \label{notat2}
\end{equation}
\begin{equation}
\vec{F}_{\alpha m}^{(j)}\left( \omega ,k_y\right) =\frac{v\mp v_0}{(2\pi
)^2\lambda _0v}\int_{-\infty }^{+\infty }dtdy\int_0^{\lambda _0v/(v\mp
v_0)}dz\vec{F}_{\alpha m}^{(j)}\left( t,\vec{r}\right) e^{-i\left( \vec{k}%
^{(\alpha )}\vec{r}-\omega t\right) }  \label{four1}
\end{equation}
and $\vec{F}_{\alpha m}^{(j)}\left( t,\vec{r}\right) $ is the corresponding
field for the $j$-th particle.

In the expansions (\ref{fields3}) the Fourier components are determined by
corresponding boundary conditions for the fields on the interface (\ref
{profile}). To reveal the structure of corresponding equations we shall
consider only one of these conditions, namely that of the equality of normal
components of the induction vector at the interface: 
\begin{equation}
\varepsilon _1\vec{n}\cdot \left[ \vec{E}_0\left( t,\vec{r}\right) +\vec{E}%
_1^{^{\prime }}\left( t,\vec{r}\right) \right] =\varepsilon _2\vec{n}\cdot 
\vec{E}_2\left( t,\vec{r}\right) ,\quad x=-d+f(\xi )\quad   \label{bound1}
\end{equation}
where 
\begin{equation}
\vec{n}=\frac{\left( 1,0,-\partial f/\partial z\right) }{\sqrt{1+\left(
\partial f/\partial z\right) ^2}}  \label{normal}
\end{equation}
is the normal to the boundary depending on $k_0z\mp \omega _0t$. Note that,
unlike to the static case considered in \cite{mkrt4}, here the boundary
conditions (\ref{bound1}) can not be directly imposed on the fields spectral
components for the reason that the normal $\vec{n}$ is time dependent. By
substituting the corresponding expressions (\ref{fields2}) and (\ref{fields3}%
) for the fields into (\ref{bound1}), multiplying by $\exp \left[
-ik_y^{\prime }y+i\omega ^{\prime }t(1\mp v_0/v)\right] $ and integrating
over $y$ and $t$ for fixed $z\mp v_0t$ one obtains 
\[
\vec{n}\cdot \sum_{m=-\infty }^{+\infty }\left\{ \varepsilon _\alpha \vec{E}%
_{\alpha m}^{(j)}\left( \omega ,k_y\right) \exp \left[ ik_x^{\left( \alpha
\right) }\left( -d+f(\xi )\right) \right] \right\} _{\alpha =1}^{\alpha
=2}e^{-2\pi imz/\lambda _0}\mid _{\omega =\omega ^{\prime }\mp m\omega _0}=
\]
\begin{equation}
=\vec{n}\cdot \varepsilon _1\vec{E}_0\left( \omega ,k_y\right) \exp \left[
-ig_0\left( -d+f(\xi )\right) \right] \exp \left( ig_0X_j-ik_yY_j-i\frac
\omega vZ_j\right) \mid _{\omega =\omega ^{\prime }}  \label{bound2}
\end{equation}
It follows from here that for the new functions 
\begin{equation}
\vec{E}_{\alpha m}^{\left( 1\right) }\left( \omega ,k_y\right) =\vec{E}%
_{\alpha m}^{\left( j\right) }\left( \omega ,k_y\right) \exp \left(
-ig_0(\omega _1)X_j+ik_yY_j+i\frac{\omega _1}vZ_j\right) ,\quad \omega
_1=\omega \pm m\omega _0  \label{fields4}
\end{equation}
the corresponding set of boundary conditions does not depend on the particle
number $j$ : 
\[
\vec{n}\cdot \sum_{m=-\infty }^{+\infty }\varepsilon _\alpha \vec{E}_{\alpha
m}^{\left( 1\right) }\left( \omega ,k_y\right) \exp \left[ ik_x^{\left(
\alpha \right) }\left( -d+f(\xi )\right) \right] _{\alpha =1}^{\alpha
=2}e^{-2\pi imz/d}\mid _{\omega =\omega ^{\prime }\mp m\omega _0}=
\]
\begin{equation}
=\varepsilon _1\vec{n}\cdot \vec{E}_0\left( \omega ,k_y\right) \exp \left[
ig_0\left( d-f(\xi )\right) \right] \mid _{\omega =\omega ^{\prime }}
\label{bound3}
\end{equation}

The values $\vec{E}_{\alpha m}^{\left( 1\right) }$ determine the field of a
single charge with coordinates $X_j=Y_j=Z_j=0$ at the initial moment $t=0$.
The other boundary conditions can be considered analogously. By using Eq. (%
\ref{disrel}) the formula for $\omega _1$ can be written as 
\begin{equation}
\omega _1=mk_0v\frac{1\mp (v_0/c)\sqrt{\varepsilon _\alpha }\cos \theta }{%
1-\beta \sqrt{\varepsilon _\alpha }\cos \theta }\approx \frac{mk_0v}{1-\beta 
\sqrt{\varepsilon _\alpha }\cos \theta }  \label{omega1}
\end{equation}
where we have used the fact that surface wave velocity is much less than $c$%
. For example, in case of surface wave excited in $SiO_2$ one has $%
v_0\approx 3\cdot 10^5cm/\sec $ and $v_0/c\approx 10^{-5}$.

As it follows from the above analysis the Fourier components of the bunch
field can be presented in the form 
\begin{equation}
\vec{F}_{\alpha m}=\sum_{j=1}^N\vec{F}_{\alpha m}^{\left( j\right) }=\vec{F}%
_{\alpha m}^{\left( 1\right) }\sum_{j=1}^N\exp \left( ig_0(\omega
_1)X_j-ik_yY_j-i\frac{\omega _1}vZ_j\right)  \label{fields5}
\end{equation}
At large distances $\rho =\sqrt{x^2+y^2}$ from the bunch trajectory we find
by means of the stationary phase method 
\begin{equation}
\vec{F}_{\alpha m}\left( \omega ,\vec{r}\right) =\sqrt{\frac{2\pi \left|
\omega \right| }{ic\rho }\sqrt{\varepsilon _\alpha }\sin \theta }\cos
\varphi \vec{F}_{\alpha m}\left( \omega ,k_y\right) e^{i\vec{k}^{(\alpha )}%
\vec{r}}  \label{fields6}
\end{equation}
\[
\vec{r}=\left( \rho \cos \varphi ,\rho \sin \varphi ,z\right) ,\quad \vec{k}%
^{\left( \alpha \right) }=\frac \omega c\sqrt{\varepsilon _\alpha }\left(
\sin \theta \cos \varphi ,\sin \theta \sin \varphi ,\cos \theta \right) 
\]

The density of radiation energy flux during the whole time of emission is
determined as 
\begin{equation}
\frac c{4\pi }\int_{-\infty }^{+\infty }dt\left[ \vec{E}\vec{H}\right]
=c\int_0^\infty d\omega \sum_mRe\left[ \vec{E}_m\left( \omega ,\vec{r}%
\right) \vec{H}_m\left( \omega ,\vec{r}\right) \right] \equiv \int_0^\infty
d\omega \sum_m\vec{P}_m^{\left( N\right) }\left( \omega \right)
\label{intens1}
\end{equation}
Using the expressions (\ref{fields6}) for the fields we find the spectral
density of the radiation energy flux in the medium $\alpha $ for given $m$
as 
\begin{equation}
\vec{P}_m^{\left( N\right) }\left( \omega \right) =2\pi \omega \varepsilon
_\alpha \sin \theta \cos ^2\varphi \left| \vec{E}_{\alpha m}\left( \omega
,k_y\right) \right| ^2\vec{k}^{\left( \alpha \right) }/k^{\left( \alpha
\right) }  \label{intens2}
\end{equation}
It follows from (\ref{fields5}) that for an arbitrary periodic function $f$
in (\ref{profile}) this expression can be written in the form 
\begin{equation}
\vec{P}_m^{\left( N\right) }\left( \omega \right) =\vec{P}_m^{\left(
1\right) }\left( \omega \right) S_N,\quad S_N=\left| \sum_{j=1}^N\exp \left(
ig_0(\omega _1)X_j-ik_yY_j-i\frac{\omega _1}vZ_j\right) \right| ^2
\label{intens3}
\end{equation}
where $\vec{P}_m^{\left( 1\right) }\left( \omega \right) $ is the
corresponding function for the radiation of a single charge with coordinates 
$x=y=z=0$ at the initial moment $t=0$. The formula for $\vec{P}_m^{\left(
1\right) }\left( \omega \right) $ is derived in \cite{mkrt1}-\cite{mkrt3}
and the dependence of the corresponding radiation intensity on various
parameters is investigated. For this reason here we will mostly concerned
with bunch form factor.

As it directly follows from Eq. (\ref{bound3}) the dependence of the
radiation intensity in the medium $\alpha $ on the distance of the electron
trajectory from the non-excited interface, $d$, is determined by the factor 
\begin{equation}
\exp \left( -\frac{2\omega }{v}dRe \sigma \right) ,\quad \sigma =\left[
\left( 1-\beta ^2\varepsilon _1\right) \frac{\omega _1^2}{\omega ^2}+\beta
^2\varepsilon _\alpha \sin ^2\theta \sin ^2\varphi \right] ^{1/2},\quad
g_0=i\frac \omega v\sigma  \label{distep}
\end{equation}
This factor does not depend on the specific form of the wave profile in (\ref
{profile}) and is determined by the dependence of the field spectral
components for uniformly moving particle on distance $d$. As mentioned above
for relativistic particles with $\beta ^2\varepsilon _1\stackrel{<}{\sim }1$
one has $\omega _1\approx \omega $ and the radiation at large azimuthal
angles is exponentially suppressed and the radiation distribution is
strongly anisotropic. For the radiation in the vacuum at $\varphi \approx 0$
the factor is equal to $\exp (-2\omega _1d/v\gamma )$, with $\gamma =1/\sqrt{%
1-\beta ^2}$ been the Lorentz factor.

In the case of $v\sqrt{\varepsilon _1}<c$, the quantity $\sigma $ is real
and the intensity exponentially decreases with increasing distance. The same
is the case for the directions of radiations satisfying the condition 
\begin{equation}
\sin ^2\theta \sin ^2\varphi >1-c^2/v^2\varepsilon _1  \label{cond4}
\end{equation}
when $v\sqrt{\varepsilon _1}>c$. For the last case and directions off the
region (\ref{cond4}) , the radiation intensity does not depend on particle
distance from a surface of the periodic structure in the absence of
absorption. This corresponds to the reflection of Cherenkov radiation
emitted in the first medium.

\section{Bunch form factor and coherent effects}

We shall assume that the coordinates of the $j$-th particle are independent
random variables and the location probability of the $j$-th particle at a
given point is independent of the particle number. By averaging the quantity
(\ref{intens3}) over the positions of a particle in the bunch and making a
summation similar to \cite{gev}, we obtain 
\begin{eqnarray}
\langle \vec{P}_m^{(N)}\rangle  &=&\langle S_N\rangle \vec{P}_m^{(1)}
\label{meanintens} \\
\langle S_N\rangle  &=&Nh+N(N-1)\left| h_xh_yh_z\right| ^2  \nonumber
\end{eqnarray}
Here we have introduced the notations 
\begin{eqnarray}
h &=&\langle \exp \left( -\frac{2\omega }vXRe\sigma \right) \rangle ,\quad
h_l=\langle \exp \left( iK_ll\right) \rangle ,\quad l=x,y,z,  \label{notat20}
\\
K_x &=&-i\frac \omega v\sigma ,\quad K_y=k_y=\frac \omega c\sqrt{\varepsilon
_0}\sin \theta \sin \varphi ,\quad K_z=\frac{\omega _1}v
\end{eqnarray}
where $\sigma $ and $\omega _1$ are given by relations (\ref{distep}) and (%
\ref{fields4}). In particular, when $\omega _0=0$ we obtain the case of
Smith-Purcell radiation, generated at the passage of bunched electrons above
the surface of a diffraction grating. The functions $\left| h_l\right| ^2$
determine bunch form factors in corresponding directions. In the right hand
side of the expression (\ref{meanintens}) the summand proportional to $N^2$
determines the contribution of coherent effects. The advantage of coherent
radiation is that we can generate intense radiation by a beam of a low
average current. The intensity of this radiation is determined by a bunch
shape. Conventionally it is assumed that the coherent radiation is produced
at wavelengths equal and longer than the electron bunch length. However as
we shall see below this conclusion depends on the distribution of electrons
in the bunch and can not be valid for a number of realistic distributions.

As it follows from (\ref{notat20}) the form factors in $y$ and $z$
directions are determined by the Fourier transforms of the corresponding
bunch distributions. First let us consider a Gaussian distribution. For this
distribution of electrons, 
\begin{equation}
f_l=\frac 1{\sqrt{2\pi }b_l}\exp \left( -\frac{l^2}{2b_l^2}\right) ,\quad
l=x,y,z  \label{gauss1}
\end{equation}
with $b_x$, $b_y$ and $b_z$ being corresponding characteristic sizes of the
bunch, we have after averaging (assuming that all particles of the bunch are
in a medium with permittivity $\varepsilon _1$ and therefore $b_x<d-a$, with 
$a$ being the surface wave amplitude) 
\begin{equation}
h=\exp \left( \frac{2\omega ^2}{v^2}(Re\sigma )^2b_x^2\right) ,\quad
\left| h_x\right| ^2=\exp \left( \frac{\omega ^2}{v^2}\sigma ^2b_x^2\right)
,\quad h_l=\exp \left( -K_l^2b_l^2/2\right) ,l=y,z  \label{factor2}
\end{equation}
By substituting these relations into (\ref{meanintens}) one obtains the
following expression of the form factor 
\begin{equation}
\langle S_N\rangle =N\exp \left( \frac{2\omega ^2}{v^2}(Re\sigma
)^2b_x^2\right) \left[ 1+(N-1)\exp \left( -\frac{\omega ^2}{v^2}\left|
\sigma \right| ^2b_x^2-k_y^2b_y^2-\frac{\omega _1^2b_z^2}{v^2}\right)
\right]   \label{factgauss}
\end{equation}
where the second summand in square brackets determines the relative
contribution of coherent effects into the radiation intensity. For
non-relativistic electrons when $b_y\stackrel{>}{\sim }b_{x,z}$ the
corresponding exponent is equal to 
\[
\exp [-(2\pi m/\lambda _0)^2(b_x^2+b_z^2)-(2\pi /\lambda )^2b_y^2\sin
^2\theta \sin ^2\varphi ]
\]
with $\lambda _0$ being surface wave wavelength (here we consider the case $%
\varepsilon _1=1$). Note that in this case for the wavelength of radiation
one has $\lambda \sim \lambda _0/\beta m$ and the coherent effects are
exponentially suppressed when $\lambda \stackrel{<}{\sim }b_l/\beta $ . For
a relativistic bunch the relative contribution of coherent effects is equal
to 
\begin{equation}
N\exp \{-(2\pi /\lambda )^2[(b_x^2+b_y^2)\sin ^2\theta \sin ^2\varphi
+b_z^2]\}  \label{relativecoh}
\end{equation}
for $\sin \theta \sin \varphi >\gamma ^{-1}$, and 
\begin{equation}
N\exp \{-(2\pi b_x/\lambda \gamma )^2-(2\pi /\lambda )^2(b_z^2+b_y^2\sin
^2\theta \sin ^2\varphi )\}  \label{relativecoh1}
\end{equation}
for the radiation with $\sin \theta \sin \varphi \stackrel{<}{\sim }\gamma
^{-1}$. As we see in this case the transverse form factor is strongly
anisotropic. It follows from (\ref{factgauss}) that for a real $\sigma $,
fixed electron number and fixed distance of the bunch's center from surface
wave the radiation intensity exponentially increases with increasing $b_x$.
This is because the number of electrons passing close to the surface wave
increases. The number of electrons with long distances will also increase.
But the contribution of close electrons surpasses the decrease of the
intensity due to distant ones.

As we see for a Gaussian distribution the relative contribution of coherent
effects is exponentially suppressed in the case $b_i>\lambda $, with $%
\lambda $ being the radiation wavelength. This result is a consequence of
the mathematical fact that in the case of a function $f(x)\in C^\infty (R)$
one has the following estimate for the integral (see, for example, \cite
{fedor}) 
\begin{equation}
F(u)\equiv \int_{-\infty }^{+\infty }f(l)e^{iul}dl=O(u^{-\infty }),\quad
u\rightarrow +\infty  \label{factornotat}
\end{equation}
where $u=2\pi b_z/\lambda $ in the case of longitudinal form factor of the
relativistic bunch.

Up to now we have considered the case of Gaussian distribution in the bunch.
However it should be noted that due to various beam manipulations the bunch
shape can be highly non-Gaussian (see, for instance, \cite{bane1}). In (\ref
{factornotat}) the continuity condition for the function $f(l)$ and infinite
number of its derivatives is essential. It can be easily seen that when $%
f(l)\in C^{n-1}(R)$, and the derivative $f^{(n)}(l)$ is discontinuous at
point $l_1$, then the asymptotic estimate 
\begin{equation}
F(u)=(-iu)^{-n-1}\left[ f^{(n)}(l_1+)-f^{(n)}(l_1-)\right] ,\quad
u\rightarrow +\infty  \label{asympdiscont}
\end{equation}
takes place. Unlike the case of (\ref{factornotat}) now the form factor for
the short wavelengths decreases more slowly, as power-law, $(2\pi
b_l/\lambda )^{-n-1}$. In the coherent part of the radiation this form
factor is multiplied by a large number, $N$, of particles per bunch and
coherent effects can dominate for 
\begin{equation}
2\pi b_l/\lambda <N^{1/2(n+1)}  \label{cohcondpower}
\end{equation}
and the radiation intensity is enhanced by the factor $N(\lambda /2\pi
b_l)^{2(n+1)}$. Since conventionally there are $10^8-10^{10}$ electrons per
bunch (see e.g. \cite{bane1}) the condition (\ref{cohcondpower}) can be
easily meet even in the case $2\pi b_l/\lambda >1$. For instance, in the
case of $n=1,N\sim 10^{10}$ coherent radiation is dominant for $100\lambda
>b_l$.

As an example we shall consider the electron distribution function having
the asymmetric Gaussian form (about the asymmetric distribution of electrons
in a bunch see, for example, \cite{bane})

\begin{equation}
f(z)=\frac 2{\sqrt{2\pi }(1+p)b_l}\left[ \exp \left( -\frac{l^2}{2p^2b_l^2}%
\right) \theta (-l)+\exp \left( -\frac{l^2}{2b_l^2}\right) \theta (l)\right]
,  \label{asgauss}
\end{equation}
where $\theta \left( l\right) $ is the unit step function, $l_0=(1+p)b_l$ is
the characteristic bunch length, parameter $p$ determines the degree of
bunch asymmetry. The corresponding function $F(u)$ can be easily found (see
also \cite{gev},\cite{gev1}) 
\begin{equation}
F(u)=\frac 1{p+1}\left\{ e^{-t^2}+pe^{-p^2t^2}-\frac{2i}{\sqrt{\pi }}\left[
W(t)-pW(pt)\right] \right\} ,  \label{factor3}
\end{equation}
where 
\begin{equation}
W(t)=\int_0^t\exp (l^2-t^2)dl,\quad t=\frac{ub_l}{\sqrt{2}}  \label{notat3}
\end{equation}
Note that the expression $\left| F(u)\right| ^2$ is invariant with respect
to the replacement $p\rightarrow 1/p$, $b_l\rightarrow b_lp$ that
corresponds to the mirror reversal of the bunch. When the electron
distribution is symmetric $(p=1)$ , the second summand in square brackets of
(\ref{factor3}) is equal to zero and, as was mentioned earlier, the form
factor exponentially decreases for the short wavelengths $\lambda <2\pi
b_l/\beta $ . For $pt\gg 1$ the asymptotic behaviour of the function $F(u)$
can be easily found from (\ref{factor3}) and has the form 
\begin{equation}
F(u)\sim i\sqrt{\frac 2\pi }\frac{1-p}{u^3b_l^3p^2}  \label{asympasymgauss}
\end{equation}
Note that this formula can be obtained also directly from (\ref{asympdiscont}%
) by taking into account that in the case of (\ref{asgauss}) one has $%
f(l)\in C^1(R)$ and the second derivative is discontinuous ($n=2$ in (\ref
{asympdiscont})).

Let us discuss in more detail the case when the electrons are normally
distributed in the $x$ and $y$ directions and the distribution function in
the $z$ direction has an asymmetric Gaussian form (\ref{asgauss}). Now the
expression (\ref{notat20}) for $\langle S_N\rangle $ can be written as 
\begin{equation}
\langle S_N\rangle =N\exp \left( \frac{2\omega ^2}{v^2}\sigma ^2b_x^2\right)
\left[ 1+(N-1)\exp \left( -\frac{\omega ^2}{v^2}\sigma
^2b_x^2-k_y^2b_y^2\right) \left| F(\omega _1/v)\right| ^2\right]
\label{esen1}
\end{equation}
where the second summand in the square brackets determines the contribution
of coherent effects into the radiation intensity. In case of relativistic
bunch and $\varphi >\gamma ^{-1}$ , the exponent in this summand is of an
order of $(2\pi b_i/\lambda )^2,i=x,y$ and when the transverse size of the
bunch is shorter than the radiation wavelength, then the relative
contribution of coherent effects is $\sim N\left| F(\omega _1/v)\right| ^2$.
For an asymmetrical bunch this contribution can be dominant even in case
when the bunch length is greater than the wavelength. Indeed, according to (%
\ref{asympasymgauss}) even for weakly asymmetrical bunch we have $N\left|
F\right| ^2\sim N(v/\omega _1b_z)^6$ and the radiation is coherent if $b_z%
\stackrel{<}{\sim }\lambda N^{1/6}/2\pi $ where we have taken into account
that for an relativistic bunch $\omega \gg m\omega _0$ and therefore $\omega
_1\approx \omega $ , as was mentioned above. In the case of bunch with $%
N\sim 10^{10}$ it follows hence that for $2\pi b_z/\lambda \stackrel{<}{\sim 
}10$ the radiation is coherent. For the radiation $\varphi \stackrel{<}{\sim 
}\gamma ^{-1}$, the exponent of the second summand in square brackets of Eq.
(\ref{esen1}) at $\varepsilon _1=1$ is of an order of $(2\pi b_x/\lambda
\gamma )^2$. It follows hence that for a relativistic bunch the radiation in
directions $\varphi \stackrel{<}{\sim }\gamma ^{-1}$ can be coherent even in
the case when the transverse size of the bunch is greater than the
wavelength. For this it is sufficient that the following conditions 
\begin{equation}
b_x\stackrel{<}{\sim }\frac{\gamma \lambda }{2\pi },\quad b_z\stackrel{<}{%
\sim }\frac{\lambda N^{1/6}}{2\pi }  \label{cond2}
\end{equation}
were met. The second of these conditions is written for weakly asymmetrical
bunches. In the case of strongly asymmetrical bunch the corresponding
conditions are much less restrictive: $b_z\stackrel{<}{\sim }\lambda
N^{1/2}/2\pi $ for $pb_z\ll \lambda $ .

Let $f(l,a)$ be a continuous distribution function depending on parameter $a$%
, and $\lim_{a\rightarrow 0}f(l,a)=f(l)$. The integral $F(u,a)$ for $f(l,a)$
uniformly converges and hence $\lim_{a\rightarrow 0}F(u,a)=F(u)$. It follows
from here that the estimate presented above is valid for continuous
functions as well if they are sufficiently close to the corresponding
discontinuous function (the corresponding derivative is sufficiently large,
see below). Aiming to illustrate this we shall consider the asymmetric
distribution function 
\begin{equation}
f(z,a_l,b_l,l_0)=\frac 1{4l_0}\left[ th\left( \frac{l+l_0}{a_l}\right)
-th\left( \frac{l-l_0}{b_l}\right) \right]  \label{asymth}
\end{equation}
For $l_0>a_l,b_l$ this function describes a rectangular bunch having
exponentially decreasing asymmetric tails with characteristic sizes $a_l$
and $b_l$. In the limit $a_l,b_l\rightarrow 0$ one obtains the rectangular
distribution with the bunch length $2l_0$. The explicit evaluation of the
expression (\ref{factornotat}) with the function (\ref{asymth}) leads to 
\begin{equation}
F(u,a_l,b_l,l_0)=\frac i{2ul_0}\left( \frac{\overline{a}_le^{-iul_0}}{\sinh 
\overline{a}_l}-\frac{\overline{b}_ze^{iul_0}}{\sinh \overline{b}_l}\right)
,\quad \overline{c}_l\equiv \pi uc_l/2  \label{factorforth}
\end{equation}
It follows from here that if $\overline{c}_l\sim 1$ then $F\sim (ul_0)^{-1}$
for $ul_0\gg 1$. In the limit $a_l,b_l\rightarrow 0$ from (\ref{factorforth}%
) one obtains the well known form factor for the rectangular distribution: 
\begin{equation}
F_{rect}(u,l_0)=\frac{\sin ul_0}{ul_0}  \label{factorrect}
\end{equation}
As we see the rectangular distribution is a good approximation for (\ref
{asymth}) if $a_l,b_l\ll \lambda $. The main contribution to (\ref
{factorforth}) comes from the bunch tails, i.e. from the parts of bunch with
large derivatives of the distribution function. This is the case for the
general case of distribution function as well: if $ul_0\gg 1$ the main
contribution comes from the parts of the bunch where $df/d(l/l_0)\stackrel{>%
}{\sim }u$ and in this case $F(u)\sim 1/u$, $u\rightarrow \infty $. This can
be generalized for higher derivatives as well: if $d^if/d(l/l_0)^i\ll u$, $%
i=1,...,n-1$, and $d^nf/d(l/l_0)^n\stackrel{>}{\sim }u$ then $F(u)\sim
(ul_0)^{-n}$.

For a relativistic bunch one has $u\sim 2\pi /\lambda $ for the form factors
in $y$ and $z$ directions and the conclusion can be formulated as follows.
If for the distribution function one has 
\begin{equation}
\lambda \frac{d^if}{d(l/l_0)^i}\ll 2\pi ,\quad i=1,...,n-1;\quad \lambda 
\frac{d^nf}{d(l/l_0)^n}\stackrel{>}{\sim }2\pi  \label{cohcondn}
\end{equation}
then the relative contribution of coherent effects into the radiation
intensity is proportional to $N(\lambda /2\pi l_0)^{2n}$ and the radiation
can be partially coherent in the case $\lambda <l_0$ but $\lambda >2\pi
l_0N^{-1/2n}$, with $l_0$ being the characteristic bunch size in
corresponding direction. In this case the main contribution into the
radiation intensity comes from the parts of the bunch with large derivatives
of the distribution function in the sense of second condition in (\ref
{cohcondn}). For example in the case of asymmetric distribution (\ref
{asgauss}) when $ub_l\gg 1$ and $pub_l<1$ the main contribution comes from
the left Gaussian tail with $l<0$. For this tail $df/d(l/b_l)\sim u/(pub_l)%
\stackrel{>}{\sim }u$, at $l\sim pb_l$ and therefore $F(u)\sim 1/(ub_l)$.
This can be seen directly from the exact relation (\ref{factor3}) as well by
using the asymptotic formula for the function $W(t)$.

As an another example of the bunch distribution let us consider the
superposition of two Gaussian functions 
\begin{equation}
f(l)=\frac{\exp (-l^2/2a_l^2)+\exp [\alpha -(l-l_1)^2/2b_l^2]}{\sqrt{2\pi }%
\left( a_l+e^\alpha b_l\right) }  \label{twogauss}
\end{equation}
where $l_1$ is the displacement of the centers, and $e^\alpha $
characterizes the relative heights. (As it have been noted in \cite{settak}
the measured spectrum of transition radiation generated by femtosecond
electron bunches agrees much better with a distribution consisting of a
short Gaussian core and a Gaussian tail than a Gaussian or rectangular ones
(see also \cite{bane1})). The corresponding form factor have the form 
\begin{equation}
F(u)=\frac{a_l\exp (-u^2a_l^2/2)+b_l\exp [\alpha +iul_1-u^2b_l^2/2]}{%
a_l+e^\alpha b_l}  \label{factortwogauss}
\end{equation}
For $\alpha <0$ and $a_l>b_l$ bunch length and the main contribution into
the number of particles are determined by the first summand in (\ref
{twogauss}). However the form factor is determined by the contribution of
the second one 
\begin{equation}
F(u)\sim \frac{b_le^\alpha }{a_l}\exp (iul_1-u^2b_l^2/2)
\label{factortwogauss1}
\end{equation}
when $ub_l\sim 1$ and $ua_l\gg 1$. As in the previous case the main
contribution into the bunch form factor is due to the parts where the
derivative of the distribution function is large ($n=1$ in (\ref{cohcondn}%
)). Now the relative contribution of coherent effects is proportional to $%
Ne^{2\alpha }(b_l/a_l)^2$ assuming that $a_l\gg \lambda $, $b_l\sim \lambda $%
.

By using the formulas presented above one can obtains the transversal and
longitudinal form factors (\ref{notat20}) for various combinations of the
considered distributions of electrons. We have to choose $u=\omega _1/v$ for
the longitudinal form factor, $u=k_y$ and $u=i\omega \sigma /v$ for the
factors corresponding to the $y$ and $x$ directions respectively. Note that
to obtain the formulas (\ref{meanintens}), (\ref{notat20}) we have assumed
that bunch moves in the media with $\varepsilon =\varepsilon _1$ and
therefore the distribution function in $x$ direction have to be zero or tend
zero sufficiently fast for $x<-d+a$ (with $a$ being the surface wave
amplitude), when the contribution of bunch tail with $x<-d+a$ is negligible, 
$f(x)\exp (-\omega \sigma x/v)\rightarrow 0$, $x\rightarrow -\infty $. This
is the case for a Gaussian bunch, and is not the case for (\ref{asymth}).

As an example of combination of various distributions let us consider the
case when the transverse distribution in the bunch have rectangular form
with characteristic sizes $x_0$ and $y_0$ in the $x$ and $y$ directions
correspondingly. By using the formulas presented above the bunch form factor
can be written in the form 
\begin{equation}
\left\langle S_N\right\rangle =N\frac{\sinh (2\omega \sigma x_0/v)}{2\omega
\sigma x_0/v}\left\{ 1+(N-1)\frac{\tanh (\omega \sigma x_0/v)}{\omega \sigma
x_0/v}\frac{\sin ^2k_yy_0}{k_y^2y_0^2}\left| h_z\right| ^2\right\}
\label{factorrect2}
\end{equation}
where $k_y=(\omega \sqrt{\varepsilon _1}/c)\sin \theta \sin \varphi $. The
longitudinal form factor $\left| h_z\right| ^2=\left| F(\omega _1/v)\right|
^2$ and is determined by relations (\ref{factor2}), (\ref{factor3}), (\ref
{factorforth}), (\ref{factorrect}), (\ref{factortwogauss}) for various
distributions. For example in the case of symmetric distribution function (%
\ref{asymth}) ($a_z=b_z$) 
\begin{equation}
\left| h_z\right| ^2=\left( \frac{\pi a_z}{2z_0}\right) ^2\left[ \frac{\sin
(\omega _1z_0/v)}{\sinh (\pi \omega _1a_z/2v)}\right] ^2  \label{longfactor2}
\end{equation}
Taking into account that the surface wave velocity is much less than the
light velocity, we arrive at the expression 
\begin{equation}
\frac{\omega _1a_z}v=\frac{2\pi ma_z}{\lambda _0(1-\beta \sqrt{\varepsilon _1%
}\cos \theta )}  \label{ksi1}
\end{equation}
with $\lambda _0$ being the wavelength of surface wave. In the case of
relativistic electrons this expression is equal to $2\pi a_z/\lambda $,
where $\lambda $ is the wavelength of the radiation.

\section{Conclusion}

The present paper is devoted to the radiation from an electron bunch of
arbitrary structure flying over the surface wave excited in plane interface
between media with different dielectric constants. The radiation intensity
at given direction and harmonic is presented in the form of product of the
corresponding quantity for a single electron and bunch form factor. In a
general case of a monoenergetic bunch this form factor is averaged by the
statistical distribution function of electrons. 

The various examples of transverse and longitudinal distribution functions
are considered, including Gaussian, asymmetric Gaussian, two Gaussian,
rectangular distribution with exponential tails and their combinations. For
fixed number of particles in the bunch and fixed distance of the bunch's
center from surface wave the radiation intensity exponentially increases
with increasing $b_x$, bunch size in direction perpendicular to the
interface. This is because the number of electrons passing close to the
surface wave increases. The number of electrons with long distances increase
as well. But the contribution of close electrons surpasses the decrease of
the intensity due to distant ones. 

It is shown that the radiation from a bunch can be partially coherent in the
range of wavelengths much shorter than the characteristic longitudinal size
of the bunch and the main contribution into the radiation intensity comes
from the parts of the bunch with large derivatives of the distribution
function in the sense of (\ref{cohcondn}). In this case for short
wavelengths the relative contribution of coherent effects decreases as power
- law instead of exponentially decreasing. The corresponding conditions for
the distribution function are specified (similar conditions for the specific
case of diffraction radiation and asymmetric Gaussian distribution were
derived in \cite{mkrt4}). The coherent effects lead to an essential increase
in the intensity of emitted radiation.

\end{document}